\documentclass{iau}
\usepackage{graphicx}
\usepackage{natbib}
\usepackage{xspace}
\usepackage{hyperref}
\usepackage{amsmath}

\newcommand{\Topcat}{\texttt{TOPCAT}\xspace}
\newcommand{\vaex}{\texttt{vaex}\xspace}
\newcommand{\Vaex}{\texttt{Vaex}\xspace}
\newcommand{\Python}{\texttt{Python}\xspace}

\newcommand{\Clang}{\texttt{C}\xspace}
\newcommand{\Lz}{L$_z$\xspace}
\newcommand{\Energy}{E\xspace}
\newcommand{\Gaia}{{\it Gaia}\xspace}

\title[Interactive visualization with vaex] 
{Interactive (statistical) visualisation and exploration of a billion objects with vaex}

\author[Maarten A. Breddels]   
{Maarten A. Breddels$^1$
}

\affiliation{$^1$Kapteyn Astronomical Institute, University of Groningen, P.O. Box 800, 9700 AV Groningen, The Netherlands \\ email: {\tt breddels@astro.rug.nl}}

\pubyear{2016}
\volume{325}  
\jname{Astroinformatics}
\editors{...}
\begin{document}

\maketitle

\begin{abstract}
With new catalogues arriving such as the \Gaia DR1, containing more than a billion objects, new methods of handling and visualizing these data volumes are needed. We show that by calculating statistics on a regular (N-dimensional) grid, visualizations of a billion objects can be done within a second on a modern desktop computer. This is achived using memory mapping of hdf5 files together with a simple binning algorithm, which are part of a \Python library called \vaex. This enables efficient exploration or large datasets interactively, making science exploration of large catalogues feasible. \Vaex is a \Python library and an application, which allows for interactive exploration and visualization. The motivation for developing \vaex is the catalogue of the \Gaia satellite, however, \vaex can also be used on SPH or N-body simulations, any other (future) catalogues such as SDSS, Pan-STARRS, LSST, etc. or other tabular data. The homepage for \vaex is \url{http://vaex.astro.rug.nl}.

\end{abstract}

\firstsection 
\section{Introduction}

\Gaia is an European Space Agency (ESA) cornerstone satellite mission, that aims to measure accurate astrometry (sky positions, parallax and proper motions) for over a billion stars in the Milky Way. Compared to the previous Hipparcos satellite \citep{Perryman1989Hipparcos}, which measured $\sim~120\,000$ parallaxes accurately, the \Gaia data is expected to revolutionize our knowlege of the Milky Way. The \Gaia satellite was launched on 19 December 2013 \citep{GaiaMission2016}, and we recently had its first data release of over a billion sources \citep{GaiaDR1SummaryBrown, GaiaDR1AstrometryLindegre}. Although not all sources have their five astrometric properties determined, the positions and G band fluxes/magnitudes \citep{GaiaDR1PhotometryLeeuwen} are available for every object in the catalogue.

\begin{figure*}[t]
  \begin{center}
    \begin{tabular}{cc}
			\includegraphics[scale=0.25]{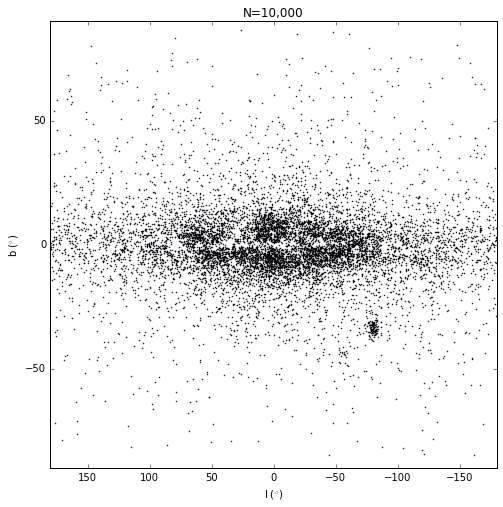} & \includegraphics[scale=0.25]{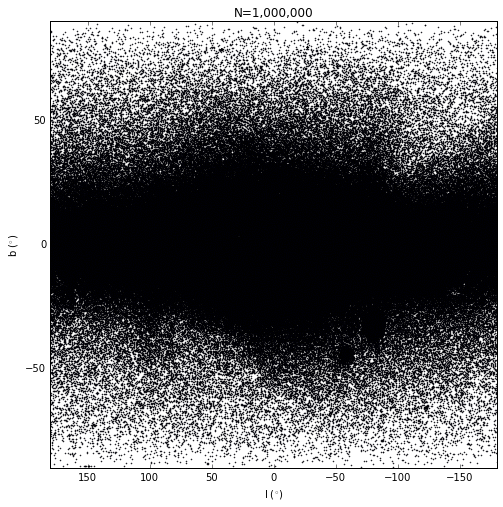}\\ \multicolumn{2}{c}{\includegraphics[scale=0.45]{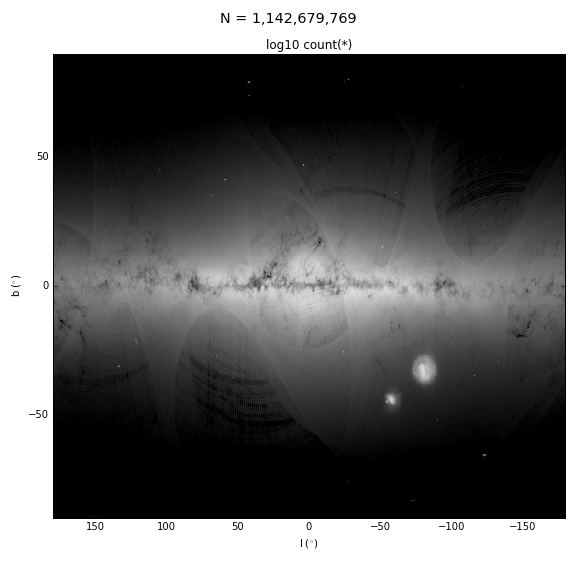}}
    \end{tabular}
    \caption{Illustration of scatter plot versus a density plot in galactic coordinates of the \Gaia DR1 catalogue showing how a scatter plot can fail, while a density plot shows the rich structure in the data. {\bf Top left: } Scatter plot showing $10\,000$ points. {\bf Top right:} Idem, with $1\,000\,000$ points. {\bf Bottom:} Density plot with $1\,142\,679\,769$ points. The top left and bottom plot show more structure in the galactic disk compred to the top right, where overplotting hides details, and the density plot shows even more structure, such as artifacts in the data due to the scanning nature of the satellite. The bottom visualization can be generated in less than a second.\label{fig:scatter_vs_density}}
    \end{center}
\end{figure*}

Working with a catalogue of a billion objects is not an easy task. However, for many science cases, as well as quality checks of the data, we would like to visualize all or large parts of the data. While scatter plots would suffice when working with the Hipparcos catalogue, doing the same for the full \Gaia catalogue, would not be useful. Apart from the long time it takes to render each individual point as a glyph, the overplotting makes the plot meaningless, as we demonstrate in Fig. \ref{fig:scatter_vs_density}. In this figure, we show how plotting a random subset of $10^4$ stars (top left panel) shows structure in the galactic disk, while plotting $10^6$ stars (top right panel) already starts to hide any structure that is present in the data, due to overplotting. In the bottom panel of this figure, we show the density of points on a grid, with low densities corresponding to black, and high densities to white. This visualization shows much more structure in the data, such as the dust that is present in the disk, as well as artifacts in the data due to the scanning nature of the satellite. Preferrably, we would like to have an interactive version of this visualization, where one would be able to zoom and pan, but also to select a region (for instance the Large Magellanic Clouds), and make other visualizations of this selections (for instance a histogram of G magnitudes).

No software package that we know of can currently handle this. While \Topcat \citep{Topcat2005ASPC} can do many of these interactive visualizations and selections, it does not scale to a billion rows. The datashader library (\url{https://github.com/bokeh/datashader}) can do the interactive visualization of a billion rows (although slower compared to our software),  but does not provide mechanisms for efficient selections and focusses on 2d.

\section{Main ideas}

\begin{figure*}[t]
  \begin{center}
    \begin{tabular}{ccc}
			\includegraphics[scale=0.32]{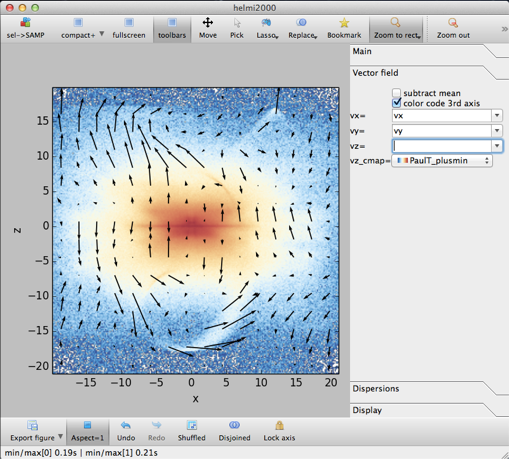} & \includegraphics[scale=0.31]{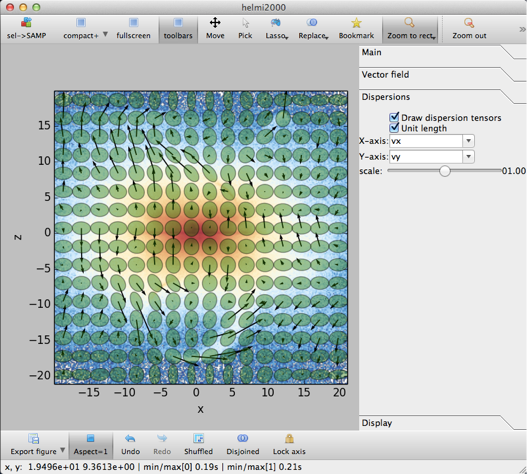} \\ \multicolumn{2}{c}{\includegraphics[scale=0.32]{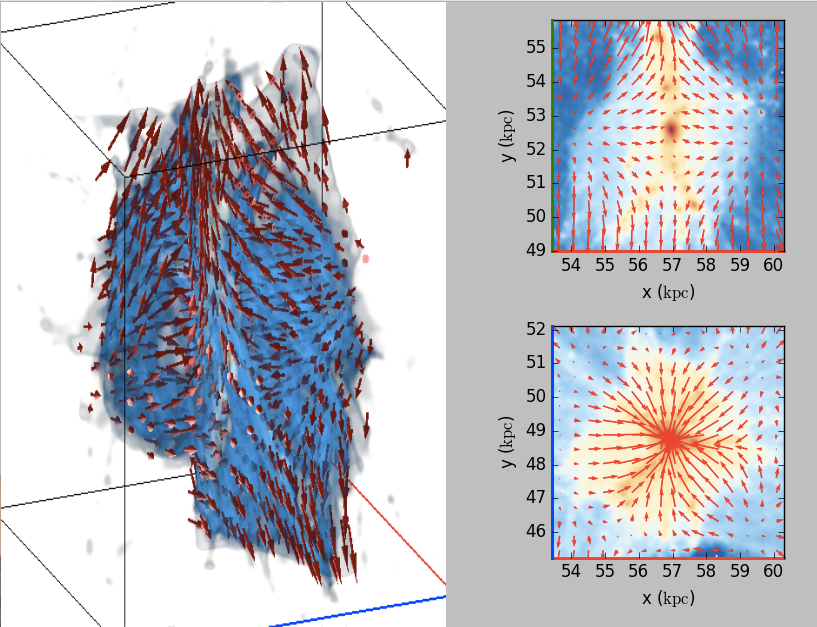}}
    \end{tabular}
    \caption{Illustration of visualization of statistics (mean velocity) using vectors (top left panel), a velocity dispersion tensor using ellipses (top right panel), or the mean velocity in 3d using vectors (bottom panel). \label{fig:vaex}}
    \end{center}
\end{figure*}

In order to do the visualization of a billion rows interactively, we would like to be able to generate a visualization as in the bottom panel of Fig. \ref{fig:scatter_vs_density} in about 1 second. If we take this as an example, where we need to process $\sim 10^9$ rows and two columns (galactic longitude l and galactic latitude b) of double precision (8 bytes per double), we need to bring a total of $10^9\times 2 \times 8~\text{bytes} = 16\times 10^9 ~\text{bytes} = 16~\text{GB} \approx 15~\text{GiB}$ of data to the CPU. With current desktop machines having a memory bandwith of $\sim 10-30$ GB/s this poses no problem.

Futhermore, with a quadcore CPU of $3$Ghz, this leaves 12 CPU cycles/row/second, which is only enough to do really simple operations. We therefore only consider doing simple statistics (counts, sums, maximum, minimum) on a regular grid. With these simple algorithms, many statistics can be calculated in the order of a second, which then can be visualized. Examples are; Calculating the counts on a regular grid (right panel of Fig. \ref{fig:scatter_vs_density}), the mean velocity represented by vectors (top left panel and bottom panel of Fig. \ref{fig:vaex}), velocity dispersion tensor (top right panel of Fig. \ref{fig:vaex}), total flux, correlation between velocity components, etc.

In order to get this performance, the data should be able to fit into main memory, otherwise the performance is limited by the storage device. To avoid unnecessary memory copies, we store the data in hdf5 files, in a column oriented way, and memory map this data.

\section{Implementation}

These ideas are implemented in a \Python package called \vaex, with the core statistical algorithms being implemented in the \Clang language. This library takes care of reading of the data, multithreading, the statistical calculations and efficient implementations of doing selections, and visualization based on matplotlib \citep{MPLHunter:2007}, or OpenGL (for the 3d rendering). The statistical algorithms work in N dimensions\footnote{Where zere dimensional would be a scalar value, such as the mean of a column.}, on either the full dataset, or a selection. Each dimension of the regular grid is defined by a column, or mathematical operations on it, the number of bins in each dimension, and the coordinates of the begin and end bin. \Vaex is open source (MIT License), the main code repository can be found on Github \footnote{\url{https://github.com/maartenbreddels/vaex}}, and its homepage is \url{http://vaex.astro.rug.nl}. Part of the \vaex software, is a Graphical User Interface (GUI) build with Qt, that enables interactive visualization (zooming and panning) and exploration (selections/queries). This last part is also distributed as a standalone software package, available for Linux and OSX. The Python package is available as source (from github), or as more easily installable pip or conda package (see the webpage for more installation details).

\section{Examples}

\begin{figure*}
  \begin{center}
    \begin{tabular}{cc}
      \includegraphics[scale=0.27]{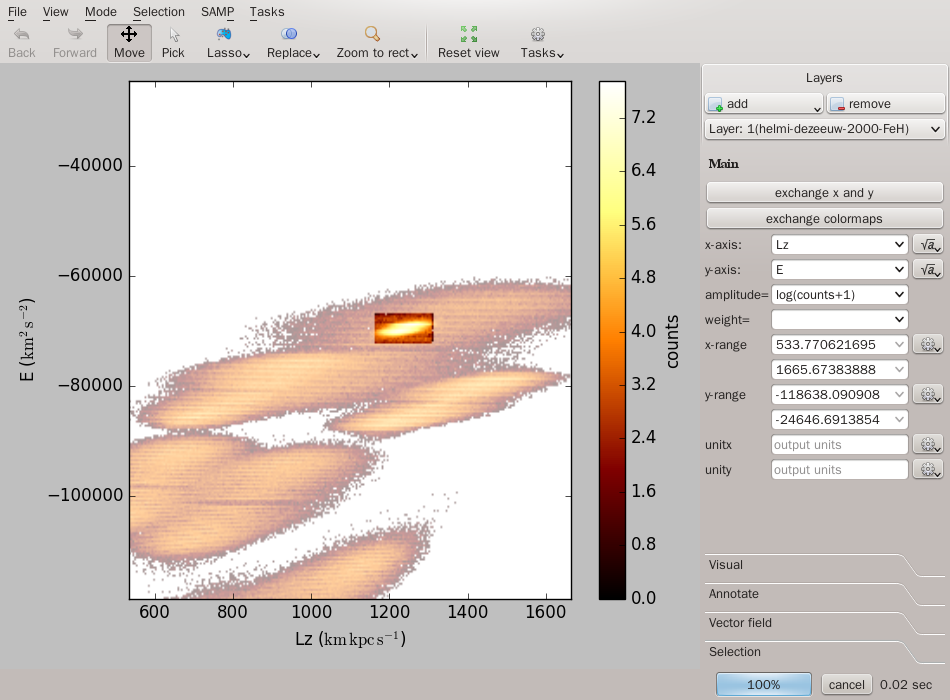} &  \includegraphics[scale=0.27]{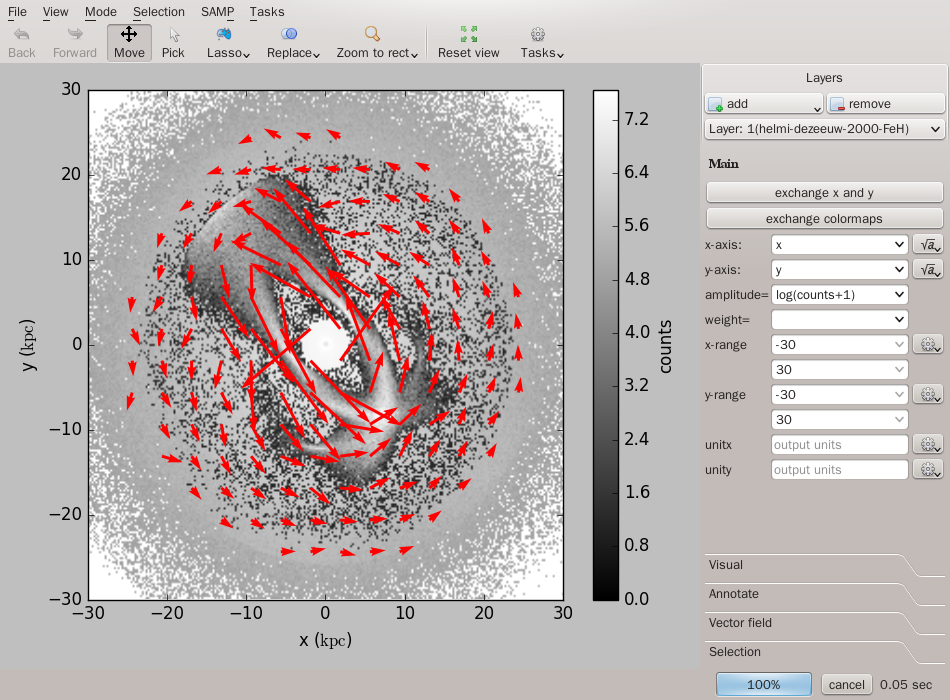}
    \end{tabular}
    \caption{{\bf Left:} 2d plot window, showing \Energy vs \Lz with a cluster in this space selected. {\bf Right:} 2d plot windows, showing $y$ vs $x$, sharing the same selection as the other window, i.e. linked views.  \label{fig:linked}}
    \end{center}
\end{figure*}

\begin{figure*}
  \begin{center}
    \begin{tabular}{c}
      \includegraphics[scale=0.32]{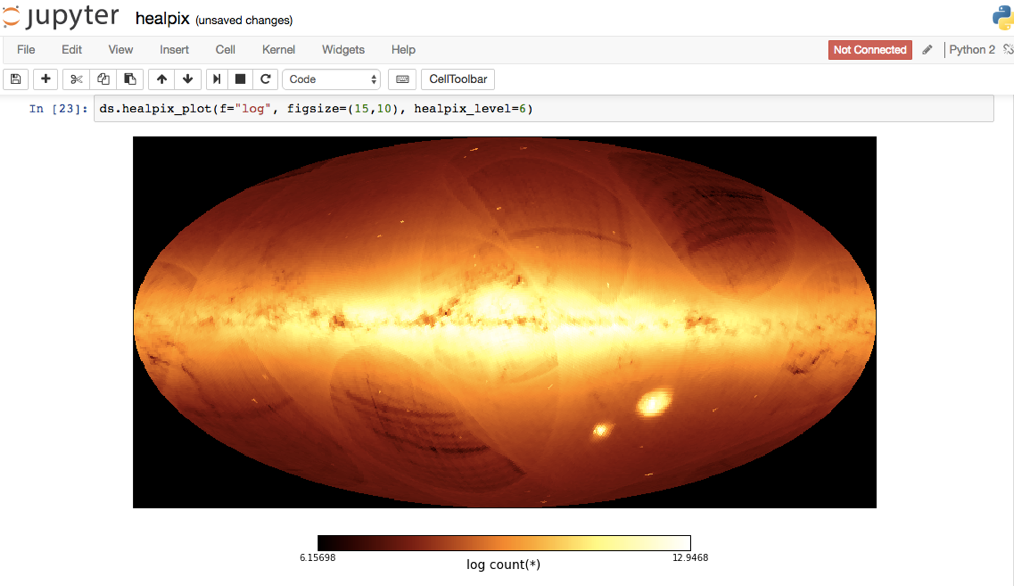} \\
			\includegraphics[scale=0.32]{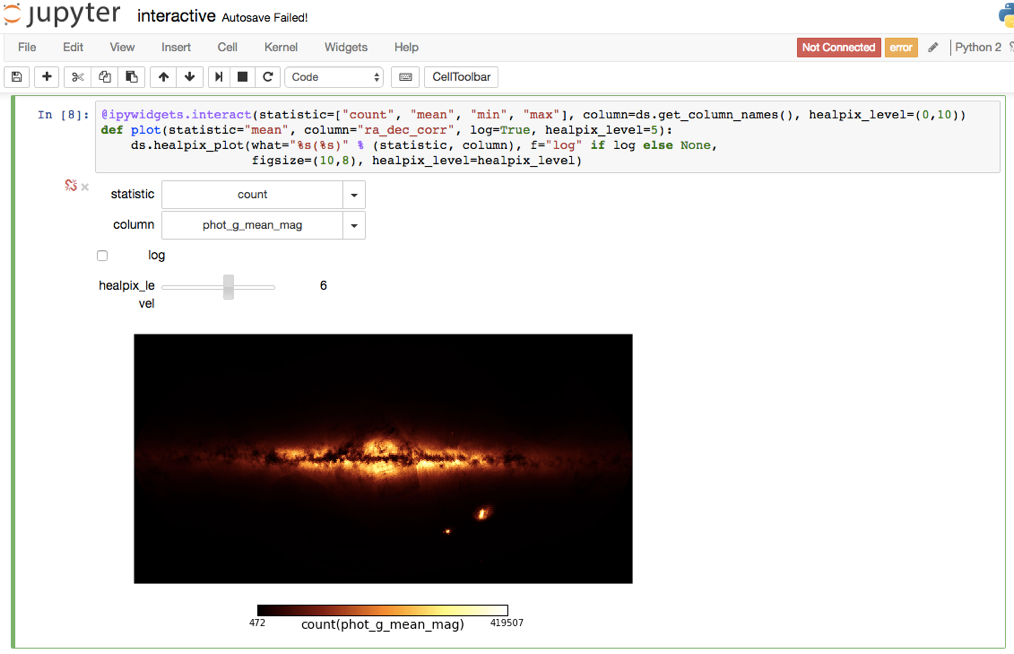}
    \end{tabular}
    \caption{{\bf Top:} Visualization using statistics on a healpix index to plot the full \Gaia DR1 catalogue sky distribution. {\bf Bottom:} Using ipywidgets, drop down menus and checkboxes can be linked to the visualization to enable custom plots with interactivity. \label{fig:notebook}}
    \end{center}
\end{figure*}

For demonstation purposes, we include a random 10\% of the data set from \citet{Helmi2000MNRAS} with the application. This dataset contains a sample of 33 simulated satellites galaxies which are disrupted in a Galactic-like halo. While most of the satellites are fully phase mixed, and not distinguisable in configuration space, there are prominent clumps present in the space of energy (\Energy) and angular momentum around the z-axis (\Lz). In Fig. \ref{fig:linked} we show the \Energy{}-\Lz{} space in the left panel, and made a selection in this space, while on the right panel we show the corresponding selection in configuration space (x-y projection) and its mean velocity using vectors. Here we see this clump in \Energy{}-\Lz{} space corresponds to a stream that is not yet fully phase mixed, and the visualization of the velocities shows its coherent space motion. This linking between the selection in multiple visualization is commonly called 'linked views'.

To visualize the billion rows of the \Gaia DR1 catalogue, we can run vaex in server mode on a machine which has enough memory to contain many columns in main memory, and connect to it using the vaex program. Now we can visualize the full \Gaia DR1 from even low end machines or laptops. For instance, the visualization shown in the bottom panel of Fig. \ref{fig:scatter_vs_density} showing the full \Gaia DR1 catalogue can be generated in less than as second.

Apart from the GUI, the \vaex library is more powerful in the Jupyter notebook \citep{PER-GRA:2007}, where the full Python programming language can be used to customize computations and visualization. In the left panel of Fig. \ref{fig:notebook} we show that in the notebook it is possible to visualize an all sky plot using statistics on a healpix index. On the right side of the same figure, we show that by using the ipywidgets\footnote{\url{https://github.com/ipython/ipywidgets}} library, with minimal effort, interactive options can be added to create custom plots.

\section{Conclusions}
The \vaex library can handle $10^9$ rows per second to calculate statistics on a (N-dimensional) regular grid. These statistics can be visualized for 1, 2 and 3 dimensions in the vaex program, and in the Jupyter notebook. The performance allows for interactive visualization (zoom and pan) and exploration (selections/queries) of massive catalogues such as \Gaia DR1. This allows one to fully exploit
not only \Gaia, but also upcoming catalogues of similar or largers such as Pan-STARRS \citep{Kaiser2010SPIE.7733E..0EK}, LSST \citep{Ivezic2008arXiv0805.2366I}, etc. or any other tabular data such as SPH or N-body simulations. Not only will \vaex enable interactive visualization and exploration of large catalogues, but the fast performance will also stimulate the exploration of ideas otherwise hampered by computational time or resources.

\begin{acknowledgements}
MB thanks Amina Helmi for making this work possible, not just financially. MB also thanks Jovan Veljanoski for his feedback on the Python API making it more human friendly. MB acknowledges financial support from NOVA. This work has made use of data from the European Space Agency (ESA) mission {\it \Gaia} (\url{http://www.cosmos.esa.int/gaia}), processed by the {\it \Gaia} Data Processing and Analysis Consortium (DPAC, \url{http://www.cosmos.esa.int/web/gaia/dpac/consortium}). Funding for the DPAC has been provided by national institutions, in particular the institutions participating in the {\it \Gaia} Multilateral Agreement.

\end{acknowledgements}

\bibliographystyle{aa}
\bibliography{iau325_breddels_vaex}

\end{document}